\documentstyle[psfig]{aipproc}
\input psfig.sty
\def\nn{\noindent}
\def\ie{{\it i.e.}}
\def\eg{{\it e.g.}}
\def\etc{{\it etc}}

\def\epem{\ifmmode e^+e^-\else $e^+e^-$\fi}
\def\to{\rightarrow}

\def\mpl{\ifmmode \overline M_{Pl}\else $\bar M_{Pl}$\fi}

\begin{document}

\rightline{\vbox{\halign{&#\hfil\cr
SLAC-PUB-8335\cr
January 2000\cr}}}
\vspace{0.8in}

\title{{$Z'$ Bosons and Kaluza-Klein Excitations at\\ Muon Colliders}
\footnote{To appear in the {\it Proceedings of the $5^{th}$ International 
Conference on Physics Potential and Development of $\mu^+ \mu^-$ Colliders}, 
Fairmont Hotel, San Francisco, CA, 15-17 December 1999}
}

\author{Thomas G Rizzo
\footnote{
E-mail:rizzo@slacvx.slac.stanford.edu. 
Work supported by the Department of Energy, 
Contract DE-AC03-76SF00515}
}

\address{Stanford Linear Accelerator Center\\
Stanford CA 94309, USA}

\maketitle

\begin{abstract}
After an extremely brief overview of the discovery reach for $Z'$ bosons, we 
will discuss the physics of Kaluza-Klein(KK) excitations of the Standard 
Model gauge bosons that can be explored by a high energy muon collider in the 
era after the LHC and TeV Linear Collider. We demonstrate that the muon 
collider is a necessary ingredient in the unraveling the properties of such KK 
states and, perhaps, proving their existence. 
\end{abstract}

\section*{Search Reaches for Z' Bosons}

The indirect search reach for new gauge bosons at future colliders has been the 
subject of much investigation but with few new results in the past couple of
years except for refinements of previously existing analyses. We refer the 
reader to the summaries provided in Ref.{\cite {snow}}.

\section*{KK Excitations of SM Gauge Bosons}

In theories with extra dimensions, $d\geq 1$, the gauge fields of the 
Standard Model(SM) will have Kaluza-Klein(KK) excitations if they are allowed 
to propagate in the bulk of the extra dimensions. If such a scenario is 
realized then, level by level, the masses of the excited states of the photon, 
$Z$, $W$ and gluon would form highly degenerate towers. The possibility that 
the masses of the lowest lying of these states could be as low as 
$\sim$ a few TeV or less leads 
to a very rich and exciting phenomenology at future and, possibly, existing 
colliders{\cite {old}}. For the case of one extra dimension compactified on 
$S^1/Z_2$ the spectrum of the excited states is given by $M_n=n/R$ and the 
couplings of the excited modes relative to the corresponding zero mode to 
states remaining on the wall at the orbifold fixed points, such as the SM 
fermions, is simply $\sqrt 2$ for all $n$. 

If such KK states exist what is the lower bound on their mass? We already know 
from direct $Z'/W'$ and dijet bump searches at the Tevatron from Run I that 
they must lie above $\simeq 0.85$ TeV. A null result for a search made with 
data from  Run II will push this limit to $\simeq 1.1$ TeV. To do 
better than this at present we must rely on the indirect effects associated 
with KK tower exchange. Such limits rely 
upon a number of additional assumptions, in particular, that the effect of KK 
exchanges is the {\it only} new physics beyond the SM.  The strongest and least 
model-dependent of these bounds arises from an analysis of charged current 
contact interactions at both HERA and the Tevatron{\cite {cornet}} 
where one obtains a bound 
of $R^{-1}>3.4$ TeV. Similar analyses have been carried out by a number of 
authors{\cite {host,rw}}; the best limit arises from an updated combined fit 
to the precision electroweak data as presented at the 1999 summer 
conferences and yields{\cite {kktest}} $R^{-1}>3.9$ TeV for the 
case of one extra dimension. 
From the previous discussion we can also draw a further conclusion for the 
case $d=1$: the lower bound $M_1>3.9$ TeV is so strong that the {\it second} KK 
excitations, whose masses must now exceed 7.8 TeV due to the above scaling 
law, will be beyond the reach of the LHC and thus the LHC will {\it at most} 
only detect the first set of KK excitations for $d=1$.

In all analyses that obtain indirect limits on $M_1$, one is actually 
constraining a dimensionless quantity such as 
\begin{equation}
V=\sum_{{\bf n}=1}^\infty {g_{\bf n}^2\over {g_0^2}} 
{M_w^2\over {M_{\bf n}^2}} \,,
\end{equation}
where, generalizing the case to $d$ additional dimensions, $g_{\bf n}$ is the 
coupling and $M_{\bf n}$ the mass of the 
$n^{th}$ KK level labelled by the set of $d$ integers {\bf n} and $M_w$ 
is the $W$ boson mass which we employ as a typical weak scale factor. For 
$d=1$ this sum is finite since $M_n=n/R$ and $g_n/g_0=\sqrt 2$ for $n>1$; 
one immediately 
obtains $V={\pi^2\over {3}}(M_w/M_1)^2$ with $M_1$ being the mass of the 
first KK excitation. From the precision data one obtains a bound on $V$ and 
then uses the above expression to obtain the corresponding bound on $M_1$. 
For $d>1$, however, independently of how the extra 
dimensions are compactified, the above sum in $V$ {\it diverges} and so it 
is not so straightforward to obtain a bound on $M_1$. We also 
recall that for $d>1$ the mass spectrum and the relative coupling strength of 
any particular KK excitation now become dependent upon how the additional 
dimensions are compactified. 

There are 
several ways one can deal with this divergence: ($i$) The simplest approach 
is to argue that as the states being summed in $V$ get heavier they approach 
the mass of the string scale, $M_s$, above which we know little and some new 
theory presumably takes over. Thus we 
should just truncate the sum at some fixed maximum value $n_{max}\simeq M_sR$ 
so that masses KK masses above $M_s$ do not contribute. 
($ii$) A second possibility is to 
note that the wall on which the SM fermions reside is not completely rigid 
having a finite tension. The authors in Ref.{\cite {wow}} argue that this wall 
tension can act like an exponential suppression of the couplings of the 
higher KK states in the tower thus rendering the summation finite, \ie, 
$g_{\bf n}^2 \to g_{\bf n}^2 e^{-(M_n/M_1)^2/n_{max}^2}$, where 
$n_{max}$ now parameterizes the strength of the exponential cut-off. 
Antoniadis{\cite {kktest}} has argued that such an gaussian suppression 
can also arise from considerations of string scattering amplitudes at 
high energies. ($iii$) A 
last scenario{\cite {schm}} is to note the possibility that the SM wall 
fermions may have a finite size in the extra dimensions which smear out and 
soften the couplings appearing in the sum to yield a finite result. In this 
case the suppression is also of the Gaussian variety. Table I shows 
how the $d=1$ lower bound of 3.9 TeV for the mass of $M_1$ changes as we 
consider different compactifications for $d>1$. We see that in some cases the 
value of $M_1$ is so large it will be beyond the mass range accessible to the 
LHC as it is for all cases of the $d=3$ example.

\vspace*{0.3cm}
\begin{table*}[htpb]
\caption{Lower bound on the mass of the first KK state in TeV resulting from 
the constraint on $V$ for the case of more than one dimension. `T'[`E'] labels 
the result 
obtained from the direct truncation (exponential suppression). Cases 
labeled by an asterisk will be observable at the LHC. $Z_2\times Z_2$ and 
$Z_{3,6}$ correspond to compactifications in the case of $d=2$ 
while $Z_2\times Z_2\times Z_2$ is for the case of $d=3$.}
\begin{tabular}{lcccccc}
     &\multicolumn{2}{c}{ $Z_2\times Z_2$}  &\multicolumn{2}{c}{$Z_{3,6}$} & 
\multicolumn{2}{c}{$Z_2\times Z_2\times Z_2$} \\ 
\tableline
\tableline
$n_{max}$ &  T   &  E   &  T   &  E   &  T   &  E    \\
\tableline
2   & 5.69$^*$  & 4.23$^*$  &6.63$^*$  &4.77$^*$ &8.65  & 8.01  \\  
3   & 6.64   & 4.87$^*$ & 7.41 &5.43$^*$ & 11.7 & 10.8 \\
4   & 7.20   & 5.28$^*$ & 7.95 &5.85$^*$ & 13.7 & 13.0 \\
5   & 7.69   & 5.58$^*$ & 8.36 &6.17$^*$ & 15.7 & 14.9 \\
10  & 8.89   & 6.42     & 9.61 &7.05  & 23.2  &  22.0  \\
20  & 9.95   & 7.16     & 10.2 &7.83  & 33.5  &  31.8  \\
50  & 11.2   & 8.04     & 12.1 &8.75  & 53.5  &  50.9  \\
\end{tabular}
\end{table*}
\vspace*{0.4cm}

\section*{SM KK States Before the Muon Collider}

Let us return to the $d=1$ case at the LHC where the degenerate KK states 
$\gamma^{(1)}$, $Z^{(1)}$, $W^{(1)}$ and $g^{(1)}$ are potentially visible. 
It has been shown {\cite {kktest}} that for masses in excess of $\simeq 4$ 
TeV the $g^{(1)}$ resonance in dijets will be washed out due to its rather 
large width and the experimental jet energy resolution available at the LHC 
detectors. Furthermore, 
$\gamma^{(1)}$ and $Z^{(1)}$ will appear as a {\it single} resonance in 
Drell-Yan that cannot be resolved and 
looking very much like a single $Z'$ as can be seen in Fig.1. Thus 
if we are lucky the LHC will observe what appears to be a degenerate $Z'/W'$. 
How can we identify these states as KK excitations when we remember that the 
rest of the members of the tower are too massive to be produced? We remind 
the reader that 
many extended electroweak models exist which predict a 
degenerate $Z'/W'$. Without further information, it would seem likely that 
this would become the most likely guess of what had been found. 
In the case of the 4 TeV resonance there is 
sufficient statistics that the KK mass will be well measured and 
one can also imagine measuring the forward-backward asymmetry, 
$A_{FB}$, if not the full 
angular distribution of the outgoing leptons, since the 
final state muon charges can be signed. However, for such a heavy resonance 
it is unlikely that much further information could be obtained about its 
couplings and other properties based on the conclusion of several years of $Z'$ 
analyses. Thus we 
will never know from LHC data alone whether the first KK resonance has been 
discovered or, instead, some extended gauge model scenario has been realized. 
To make further progress we need a lepton collider.

It is well-known that future $e^+e^-$ linear colliders(LC) operating in the 
center of mass energy range $\sqrt s=0.5-1.5$ TeV will be sensitive to indirect 
effects arising from the exchange of new $Z'$ bosons with masses typically 6-7 
times greater than $\sqrt s${\cite {snow}}. This sensitivity is even greater 
in the case of KK excitations since towers of both $\gamma$ and $Z$ exist 
all of which have couplings larger than their SM zero modes. Furthermore, 
analyses have shown 
that with enough statistics the couplings of the new $Z'$ to the SM fermions 
can be extracted{\cite {snow}} in a rather precise manner, especially when 
the $Z'$ mass is already approximately known from elsewhere, \eg, the LHC. 
In the present situation, we imagine that the LHC has 
discovered and determined the mass of a $Z'$-like resonance in the 4-6 TeV 
range. Can the LC tell us anything about the couplings of this object? 

We find that it is sufficient for 
our arguments below to do this solely for the leptonic channels. The idea is 
the following: we measure the deviations in the differential cross sections 
and angular dependent Left-Right polarization asymmetry, $A_{LR}^\ell$,
for the three lepton generations and combine those with $\tau$ polarization 
data. Assuming lepton universality(which would be observed in the LHC data 
anyway), that the resonance mass is well 
determined, and that the resonance is an ordinary $Z'$ we perform a fit to 
the hypothetical $Z'$ coupling to leptons. To be specific, let us 
consider the case of only one extra dimension with a 
4 TeV KK excitation and employ a $\sqrt s=500$ GeV 
collider with an integrated luminosity of 200 $fb^{-1}$. The result of 
performing this fit demonstrate, as shown in Ref.{\cite {new}}, 
that the coupling values are 
`well determined', \ie, the size of the $95\%$ CL allowed region we find is 
quite small as we would have expected from previous $Z'$ analyses.

%
\nn
\begin{figure}[htbp]
\centerline{
\psfig{figure=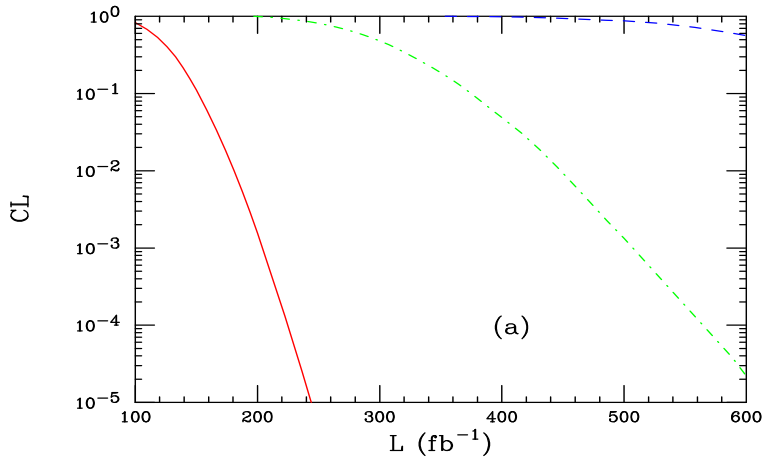,height=6.0cm,width=10cm,angle=90}}
\vspace*{1mm}
\caption{CL as a function of the integrated luminosity 
resulting from the coupling fits following from the analysis 
discussed in the text for both a 500 GeV $e^+e^-$ collider. 
In the solid(dash-dotted,dotted) curve corresponds to a first KK 
excitation mass of 4(5,6) TeV.}
\end{figure}
\vspace*{0.4mm}

The only problem with the fit is that 
the $\chi^2$ is very large leading to a very small confidence level, \ie, 
$\chi^2/d.o.f.=95.06/58$ or CL=$1.55\times 10^{-3}$! 
For an ordinary $Z'$ it has 
been shown that fits of much higher quality, based on confidence level values, 
are obtained by this same procedure. Fig.2 shows 
the results for the CL following the above approach as we vary both the 
luminosity and the mass of the first KK excitation at a 500 GeV 
$e^+e^-$ linear collider. From this analysis one finds that the resulting CL 
is below $\simeq 10^{-3}$ for a first KK excitation with a mass of 4(5,6) 
TeV when the 
integrated luminosity at the 500 GeV collider is 200(500,900)$fb^{-1}$ whereas 
at a 1 TeV  for excitation masses of 5(6,7) TeV we require luminosities of 
150(300,500)$fb^{-1}$ to realize this same CL. Barring some unknown systematic 
effect the only conclusion that one could draw from such bad fits is that the 
hypothesis of a single $Z'$, and the existence of no other new physics, 
is simply {\it wrong}.  
If no other exotic states are observed below the first KK mass at the LHC, 
this 
result would give very strong indirect evidence that something more unusual 
than a conventional $Z'$ had been found but it {\it cannot} prove that this 
is a KK state. 

%
\nn
\begin{figure}[htbp]
\centerline{
\psfig{figure=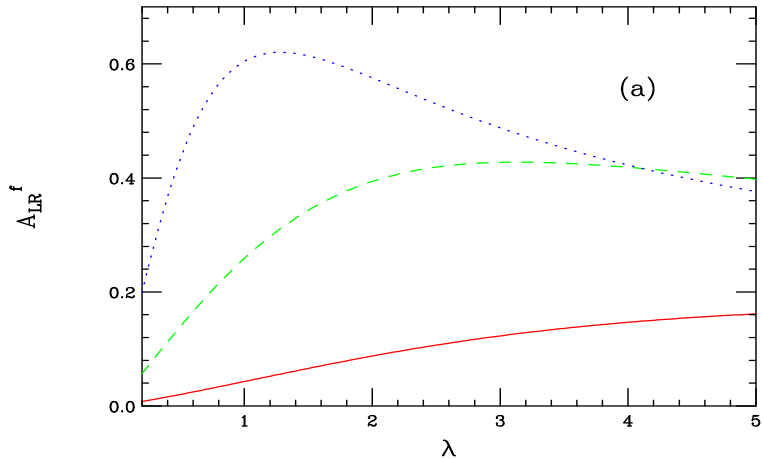,height=6.0cm,width=10cm,angle=90}}
\vspace*{0.10cm}
\caption{$A_{LR}^f$ as a function of the parameter $\lambda$ for 
$f=\ell$(solid), $f=c$(dashed) and $f=b$(dots).}
\end{figure}
\vspace*{0.4mm}

\section*{SM KK States at Muon Colliders}

In order to be completely sure of the nature of the first KK excitation, we 
must produce it directly at a higher energy lepton collider and sit on and 
near the peak of the KK resonance. To reach this mass range will most likely 
require a Muon Collider.
Sitting on the resonance there are a very large number of quantities that can 
be measured: the mass and apparent 
total width, the peak cross section, various partial 
widths and asymmetries \etc. From the $Z$-pole studies at SLC and LEP, we 
recall a few important tree-level results which we would expect to apply 
here as well 
provided our resonance is a simple $Z'$. First, we know that the value of 
$A_{LR}$, as measured on the $Z$ by SLD, does not 
depend on the fermion flavor of the final state and second, that the 
relationship $A_{LR}\cdot A_{FB}^{pol}(f)=A_{FB}^f$ holds, where 
$A_{FB}^{pol}(f)$ is the polarized Forward-Backward asymmetry as measured for 
the $Z$ at SLC and $A_{FB}^f$ is the usual Forward-Backward asymmetry. The 
above relation is seen to be trivially satisfied on the $Z$(or on a $Z'$) since 
$A_{FB}^{pol}(f)={3\over 4}A_f$, $A_{LR}=A_e$, and 
$A_{FB}^f={3\over 4}A_eA_f$. Both of these 
relations are easily shown to fail in the present case of a `dual' resonance 
though they will hold if only one particle is resonating. 

%
\nn
\begin{figure}[htbp]
\centerline{
\psfig{figure=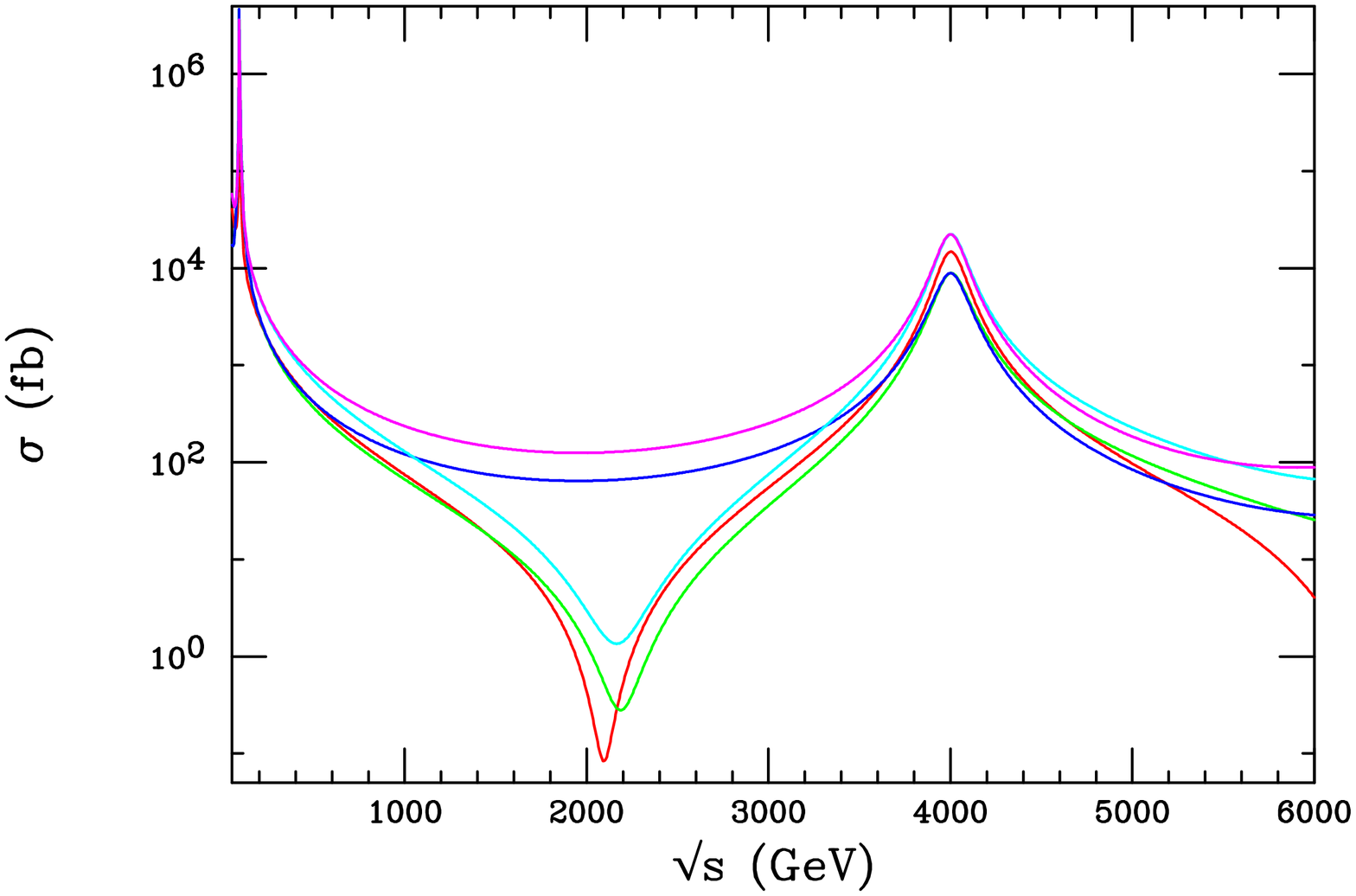,height=6.0cm,width=10cm,angle=90}}
\vspace*{9mm}
\centerline{
\psfig{figure=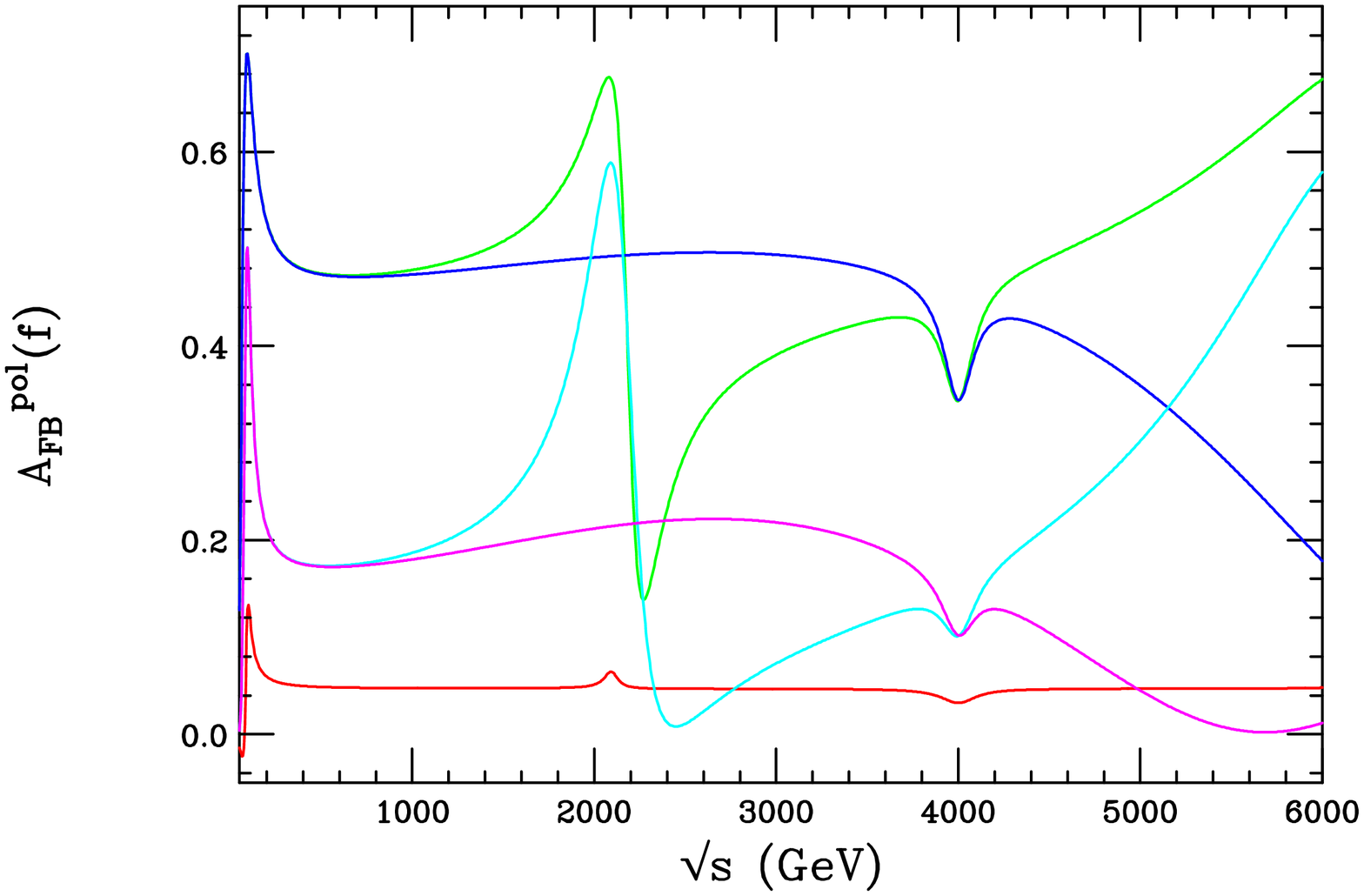,height=6.0cm,width=10cm,angle=90}}
\vspace*{0.0cm}
\caption[*]{Cross sections and polarized $A_{FB}$ for $\mu^+\mu^-\to e^+e^-$
$b\bar b$ and $c\bar c$ as functions of energy in both the 
`conventional' scenario and that of Arkani-Hamed and 
Schmaltz(AS){\cite {schm}} where the quarks and leptons are separated in 
the extra dimension by a distance $D=\pi R$. The red curve 
applies for the $\mu$ final state in either model whereas the green(blue) and 
cyan(magenta) curves label the $b$ and $c$ final states for the 
`conventional'(AS) scenario.}
\end{figure}
\vspace*{0.4mm}

A short exercise{\cite {kktest}} yields the results in Fig.2 explicitly 
showing the flavor dependence of $A_{LR}$. 
In principle, to be as model independent as possible in a numerical analysis, 
we should allow the individual widths $\Gamma_i$ of the two resonances to be 
greater than or equal to their SM 
values as such heavy KK states may decay to SM SUSY partners as well as to 
presently unknown exotic states. Since the expressions above only depend upon 
the ratio of widths, we let $R=\lambda R_0$ where $R_0$ is the width ratio  
obtained assuming that the KK states have only SM decay 
modes. We then treat $\lambda$ as a free parameter in what follows 
and explore the range $1/5 \leq \lambda \leq 5$. Once $\lambda$ is determined 
from the value of one observable all of the electroweak parameters of the 
dual resonance are completely fixed and can directly compared with data proving 
that a composite resonance corresponding to the first KK excitation 
has been discovered{\cite {new}}.

%
\nn
\begin{figure}[htbp]
\centerline{
\psfig{figure=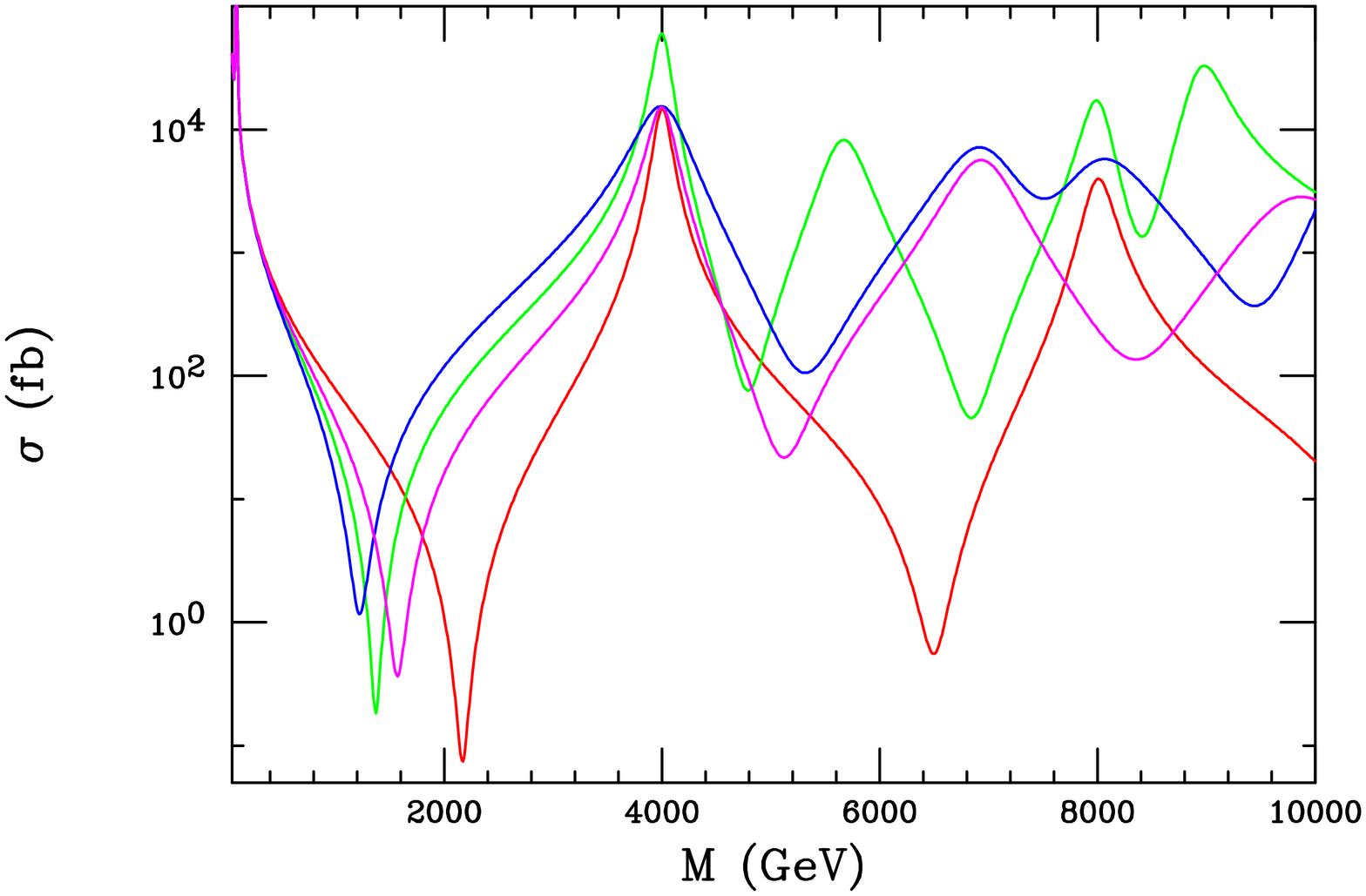,height=6.0cm,width=10cm,angle=90}}
\vspace*{0.0cm}
\caption{Same as Fig. 3a for the process $\mu^+\mu^-\to e^+e^-$ but now also 
including the models listed in Table 1 with $d=2$ assuming $M_1=4$ TeV. 
The red(green,blue,purple) curve corresponds to the $S^1/Z_2$($Z_2\times Z_2$, 
$Z_{3,6}$, $S^2$) compactifications.}
\end{figure}
\vspace*{0.4mm}

In Figs. 3a and 3b we show that although on-resonance measurements of the 
electroweak 
observables, being quadratic in the $Z^{(1)}$ and $\gamma^{(1)}$ couplings, 
will not distinguish between the usual KK scenario and that of the 
Arkani-Hamed and Schmaltz(AS) (whose KK couplings to quarks are of opposite 
sign from the conventional assignments for odd KK levels since quarks and 
leptons are assumed to be separated by a distance $D=\pi R$ in their 
scenario) the data below the peak in the hadronic channel 
will easily allow such a separation. The cross section and asymmetries for 
$\mu^+\mu^-\to e^+e^-$ is, of course, the same in both cases. 
Such data can be collected by using radiative returns if sufficient luminosity 
is available. The combination 
of on and near resonance measurements will thus completely determine the 
nature of the resonance as well as the separation between various fermions on 
the wall. Fig.4 shows that with even larger energies muon 
colliders will be able to probe both the number of extra dimensions as well as 
the geometry of their compactification manifolds since these can be uniquely 
determined by the KK excitation spectrum. 

%
\def\MPL #1 #2 #3 {Mod. Phys. Lett. {\bf#1},\ #2 (#3)}
\def\NPB #1 #2 #3 {Nucl. Phys. {\bf#1},\ #2 (#3)}
\def\PLB #1 #2 #3 {Phys. Lett. {\bf#1},\ #2 (#3)}
\def\PR #1 #2 #3 {Phys. Rep. {\bf#1},\ #2 (#3)}
\def\PRD #1 #2 #3 {Phys. Rev. {\bf#1},\ #2 (#3)}
\def\PRL #1 #2 #3 {Phys. Rev. Lett. {\bf#1},\ #2 (#3)}
\def\RMP #1 #2 #3 {Rev. Mod. Phys. {\bf#1},\ #2 (#3)}
\def\NIM #1 #2 #3 {Nuc. Inst. Meth. {\bf#1},\ #2 (#3)}
\def\ZPC #1 #2 #3 {Z. Phys. {\bf#1},\ #2 (#3)}
\def\EJPC #1 #2 #3 {E. Phys. J. {\bf#1},\ #2 (#3)}
\def\IJMP #1 #2 #3 {Int. J. Mod. Phys. {\bf#1},\ #2 (#3)}

\end{document}